\documentclass[a4paper,11pt]{article}
\pdfoutput=1 

\usepackage{jcappub} 

\usepackage[T1]{fontenc} 
\usepackage{booktabs} 
\usepackage{ dsfont }
\usepackage{float}

\usepackage{calrsfs}
\DeclareMathAlphabet{\pazocal}{OMS}{zplm}{m}{n}
\usepackage{xcolor}
\usepackage{pict2e}
\usepackage[percent]{overpic}
\usepackage{ upgreek }
\usepackage{xfrac}
\usepackage{abraces}

\title{A dark matter telescope probing the 6 to 60 GHz band}

\author[a,b]{Javier De Miguel}

\affiliation[a]{Instituto de Astrof\'isica de Canarias,\\E-38200 La Laguna, Tenerife, Spain}
\affiliation[b]{Departamento de Astrof\'isica, Universidad de La Laguna,\\E-38206 La Laguna, Tenerife, Spain}

\emailAdd{jmiguel@iac.es}

\abstract{In this article we present the Dark-photons$\&$Axion-Like particles Interferometer (DALI), 
a novel experiment designed for the detection of photon-mixing cold dark matter in the microwave 
band between 6 and 60 GHz. DALI is a haloscope for the simultaneous search for axions, axion-like particles and dark 
photons, with a number of novelties that make it unique. First, it is a dark matter telescope, with a
capacity for pointing, tracking and rastering objects and areas in the sky. This potentially 
allows one to detect relativistic dark matter particles, substructures and flows, without 
compromising the simultaneous scanning for dark matter relic particles present in the laboratory. Second, it has been designed using commercial technology. This will allow feasible manufacture 
at a reasonable cost, thereby mitigating the need for R\&D and facilitating maintenance. Finally, 
it benefits from a high sensitivity over a broad band of frequencies with only minimal reconfiguration.}

\keywords{dark matter experiments, dark matter detectors, axions, gravitational waves / experiments}

\begin{document}
\maketitle
\flushbottom 

\section{Introduction}
\label{sec:Theory}
One of the most important challenges of modern physics is to unveil the exact nature and properties of
dark matter (DM). One of the leading hypotheses is that DM is composed of a new type of 
scalar particle, generically referred as the \textit{axion}. The axion is a hypothetical pseudo-scalar 
Goldstone boson theorized by Weinberg and Wilczek (\cite{PhysRevLett.40.223}, \cite{PhysRevLett.40.279}) 
as consequence of the dynamic solution to the \textit{strong CP symmetry problem}\footnote{Charge 
and Parity or Charge-conjugation Parity symmetry, i.e., charge conjugation symmetry (C) and parity 
symmetry (P).} proposed by Peccei and Quinn (PQ) \cite{PhysRevLett.38.1440} and is predicted in multiple 
extensions of the Standard Model (SM) of Particle Physics. A fundamental parameter of the SM, the 
$\theta$ term, governs the value of the electric dipole moment of the neutron, and its absolute 
upper limit has been measured showing that $\theta$ is extremely fine tuned ($|\theta | < 10^{-11}$). In the PQ solution, $\theta$ is not a parameter but a dynamic field  
(the QCD axion\footnote{QCD stands for Quantum Chromodynamics.}), whose potential has a minimum 
to which it evolves once the Universe cools sufficiently. The axion has a light mass, induced by 
its interactions with SM particles and scales inversely to a typical energy called the PQ 
scale, $f_a$. On the other hand, numerous extensions of the SM and String theory predict a set of particles 
similar to axion, the so-called \textit{axion-like particles} or ALPs. The fundamentals of both QCD axion and ALPs have been adequately reviewed by several authors and the 
bibliography is very extensive (e.g., \cite{PhysRevD.98.030001}, \cite{MARSH20161}). Throughout this
 section, we focus only on those aspects which are relevant for the understanding of the present experiment, 
starting with the mechanism of interaction between axions and ALPs with ordinary photons. The axion-to-photon coupling Lagrangian density is 

\begin{equation}
\pazocal{L}= -\frac{1}{4} F_{\mu \nu} F^{\mu \nu}
+\frac{1}{2} \partial_{\mu} \,a\, \partial^{\mu} \,a\,
-\frac{1}{2} m_{a}^{2} \,a^2 
-\frac{\mathrm{g}_{a\gamma}}{4} F_{\mu \nu} \tilde{F}^{\mu \nu}\,a\,
-J^{\mu} A_{\mu}
\,,
\label{Eq.0}
\end{equation}
where $F^{\mu \nu}$ is the field strength tensor and $\tilde{F}$ is the dual field strength, 
$\mathord{\mathrm{g}}_{a\gamma}$ is the axion-to-photon coupling constant, $a$ the axion field, $m_a$ 
the axion mass, $J$ is density of current and the SM photon field is $A^{\mu}$.

\begin{figure}[ht!]
		\includegraphics[width=0.75\textwidth]{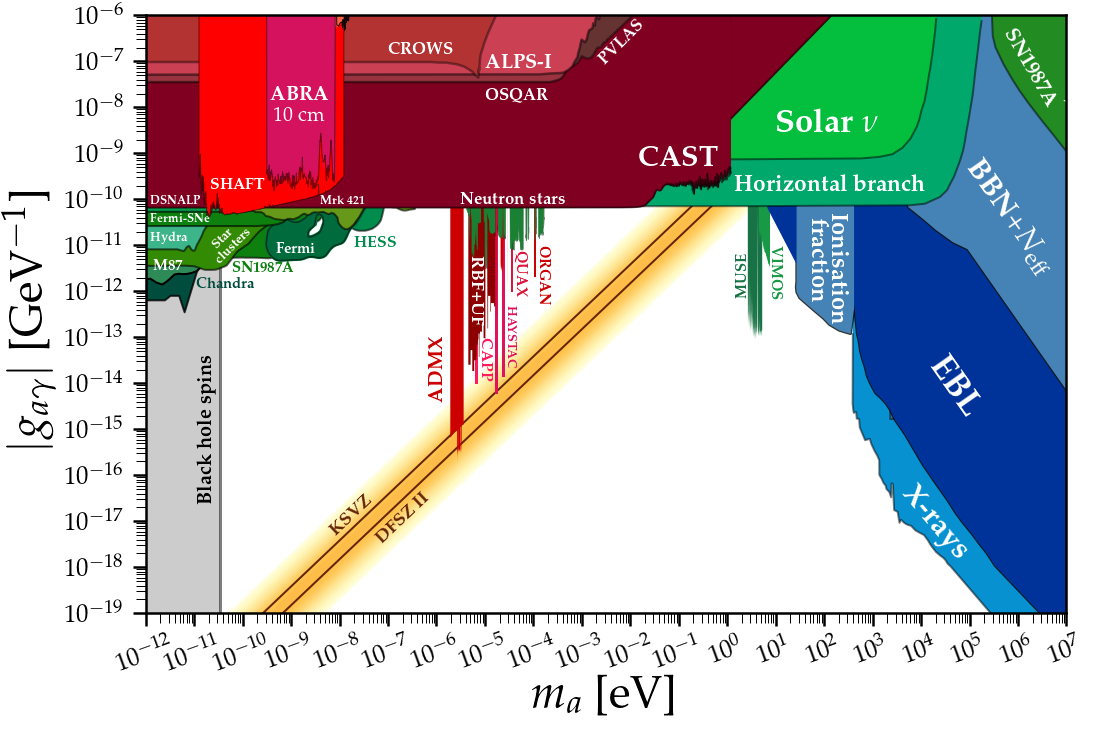}
		\centering 
		\caption{Exclusion graph of axion mass (log-scale). Laboratory experiments partially overlapping helioscopes are shaded in reddish colors. Greenish 
exclusion sectors are established by stellar hints and astronomical observations. Zones shaded in bluish colors correspond cosmology. Haloscopes are also represented in red \cite{Ciaran}.}
		\label{fig_5}
	\end{figure}
	
A set of astronomical observations and laboratory experiments constrain the range of values in which the axion and ALPs mass can exist (e.g., 
\cite{ABBOTT1983133}, \cite{DINE1983137}, \cite{PRESKILL1983127}). For  $f_a$ scales of the order of 
$10^{12}$ GeV, or masses of the order of the $\mathord{\mathrm{\mu eV}}$, the axion and ALPs are well-grounded candidates for cold dark matter (CDM). This is represented in Fig. \ref{fig_5} where the (upper) reddish zones correspond to laboratory results, including \textit{Light shining through walls} (LSW) experiments \cite{OSQAR}, measurement of axion-field induced dichroism and birefringence and vacuum magnetic birefringence (VMB) \cite{DellaValle:2015xxa} and \textit{Fifth force} experiments \cite{PhysRevD.86.015001} partially overlapping \textit{helioscopes} (\cite{PhysRevLett.69.2333}, \cite{Anastassopoulos:2017ftl}, \cite{PhysRevLett.118.261301}) and bounds depending on interactions (i.e., Primakoff \cite{PhysRev.81.899} axion loss rate, atomic axio-recombination 
or deexcitaction, axion bremsstrahlung and Compton scattering interactions) of the axion to fermions and photons in the plasma of stars, producing extra cooling (\cite{doi:10.1146/annurev.nucl.49.1.163}, \cite{Vinyoles_2015}), in green. The analysis includes observations of horizontal branch (HB) stars \cite{Straniero:2015nvc}, red giants (RGs) \cite{PhysRevLett.111.231301}, white dwarfs (WDs) \cite{Bertolami_2014}, observations of the neutrino flux duration 
from supernova SN1987A compared to numerical simulations (\cite{PhysRevD.94.085012}, \cite{Carenza_2019}), the neutron star (NS) in the SN remmant Cassiopeia A (\cite{Leinson_2014}, \cite{Keller:2012yr}, \cite{Sedrakian:2018kdm}, \cite{Giannotti_2017}) and  superradiance because of the formation of gravitationally bound states in black holes (BH)
\cite{PhysRevD.81.123530}. The search for axion in the range $10^{-6}\lesssim m_a \lesssim 10^{-3}$ is well-motivated \cite{PhysRevD.98.030001}. Haloscopes are the most sensitive detectors within this range. Thus, the most sensitive reddish regions in Fig. \ref{fig_5} represent exclusion zones from haloscopes (\cite{PhysRevLett.120.151301}, \cite{PhysRevLett.59.839}, \cite{PhysRevD.40.3153}, \cite{PhysRevD.97.092001}, \cite{Boutan:2018uoc}, \cite{Lee:2020cfj},  \cite{MCALLISTER201767}, \cite{Alesini:2019ajt}). Cosmological arguments on cosmic microwave background (CMB), X and $\gamma$ rays, Big Bang nucleonsynthesis (BBN), cosmological extragalactic background light (EBL) and $\chi_{ion}$ considerations \cite{Arias_2012} are contained in the bluish sectors. Regarding  the lower limit for axion mass, several authors suggest that ultra-light axions should not exist below $m_a\gtrsim 10^{-21}-10^{-22}$ eV (e.g., \cite{10.1093/mnras/stv624}, \cite{10.1093/mnras/stt2079}). \newline

The coupling rate of the QCD axion contains a factor derived from the model-dependent ratio of 
electromagnetic (EM) and color anomalies ($\pazocal{E}/\pazocal{C}$) given by
 $C_{a\gamma}=1.92(4)- \pazocal{E}/\pazocal{C}$,\footnote{Giving the number in 
brackets account on the uncertainty.} being $C_{a\gamma}=-\frac{\upalpha}{2\pi}\mathord{\mathrm{g}}_{a\gamma} f_a $.\footnote{Where 
$\upalpha =e^2/4\pi$ is the dimensionless \textit{fine structure constant}, $e$ being the \textit{elementary charge}.}
 The color anomaly is an integer, and is also referred  to in cosmology as the \textit{domain wall number} 
(i.e., $\pazocal{C}\equiv \pazocal{N}_{DW}$). In the case of the KSVZ 
(Kim--Shifman--Vainshtein--Zakharov) model (\cite{PhysRevLett.43.103}, \cite{SHIFMAN1980493}) 
$\pazocal{E}/\pazocal{N}_{DW}=0$ and $\pazocal{N}_{DW}=1$ are adopted. $\pazocal{N}_{DW}=1$ avoids topological issues. Topological defects would 
have catastrophic cosmological effects if stable, so the KSVZ model is consistent. The DFSZ
(Dine--Fischler--Srednicki--Zhitnitsky) model (\cite{DINE1981199}, \cite{Zhitnitsky:1980tq}) 
sets $\pazocal{E}/\pazocal{N}_{DW}$ equal 8/3 or 2/3 and $\pazocal{N}_{DW}$ equal to 6 or 3.\footnote{See DFSZI and DFSZII models.} 
Alternative mechanisms to solve topology defects in DFSZ model have been proposed (e.g., 
\cite{PhysRevLett.113.241301}). 

The \textit{PQ-phase transition} is the moment in the history of the early universe in which 
the axion angular field acquires propagating degrees of freedom. In the pre-inflationary scenario an axion mass of the order of $m_a\lesssim 20 \, \mu$eV is preferred. In the post-inflationary scenario different patches can have 
different misalignment angle $\theta_i$. These patches would remain causally disconnected and could give rise to the 
existence of DM substructures. This scenario undergoes the formation of cosmic strings and domain walls,
 solved when $\pazocal{N}_{DW}=1$, as was mentioned above. 
Assuming that all (or a significant part of) the DM in the Universe has an axionic nature, and a DM halo density of the order of $\rho_{DM}\sim 300-450\, \mathord{\mathrm{MeV\,cm^{-3}}}$ 
\cite{Gates__1995}, which will be adopted throughout the text, the value of the axion mass in this 
scenario is constrained in the range $26 \, \mu eV \lesssim m_a \lesssim  1 \,  
\mathord{\mathrm{ meV }}$. \newline 

\begin{figure}[ht!]
		\includegraphics[width=0.65\textwidth]{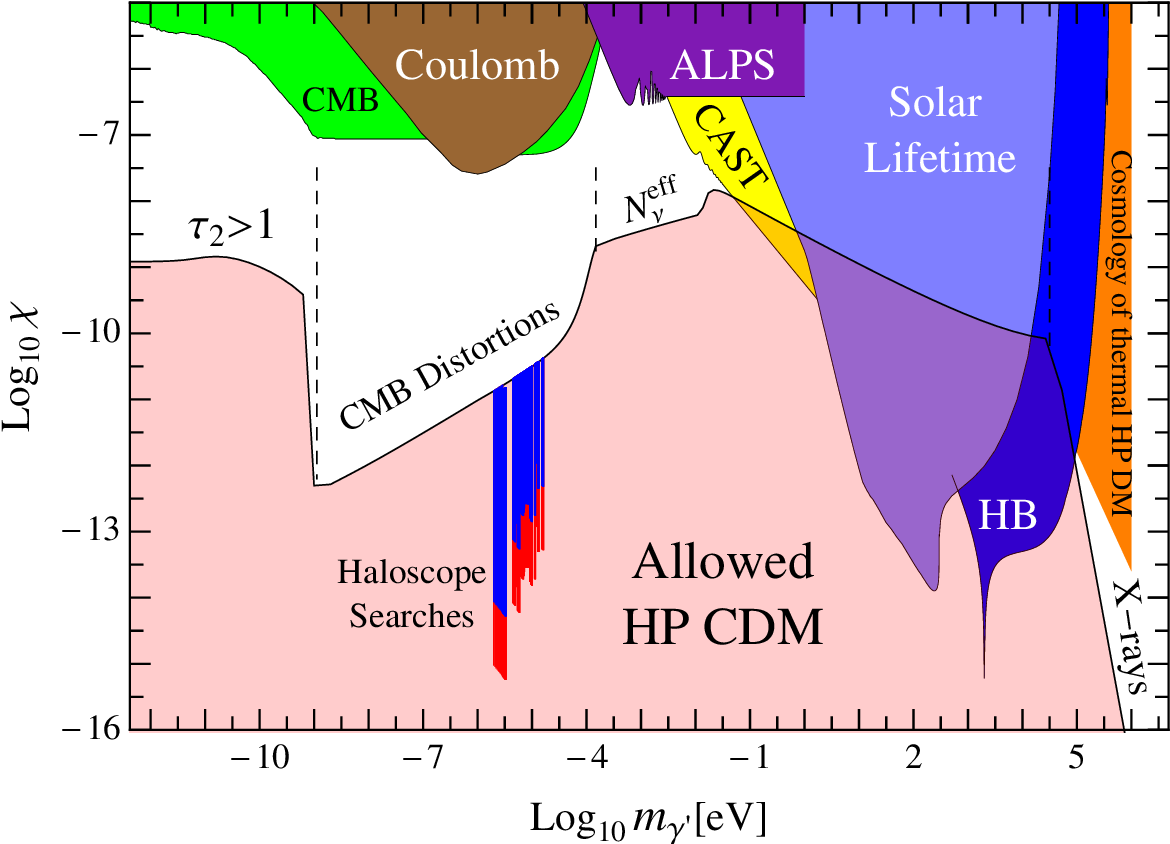}
		\centering 
		\caption{Exclusion graph for dark photon mass. Several experiments, astronomical bounds and cosmological considerations are included. Reprinted with permission from \cite{Arias_2012}.}
		\label{fig_8}
	\end{figure}
\setlength{\belowdisplayskip}{1pt}

Dark photon (DP), also called hidden photon or paraphoton (\cite{Okun}, 
\cite{Vilenkin}), arise in extensions of the SM. DPs naturally mix with SM photons. In the Lagrangian in Eq. \ref{Eq.9} we denote by $\tilde{X}^{\mu \nu}$ the field
 strength tensor of the DP field ($\tilde{X}^{\mu }$), and $F^{\nu \mu }$ the field strength tensor of the ordinary
 SM photon field ($A^{\mu}$) \cite{Horns_2013}.

\begin{equation}
\pazocal{L}= -{\frac {1}{4}}F_{\mu \nu } F^{\mu \nu }
-{\frac {1}{4}}\tilde{X}_{\mu \nu }\tilde{X}^{\mu \nu }
-{\frac {\chi}{2}}F_{\mu {\nu} } \tilde{X}^{\mu \nu }
+{\frac {1}{2}}m_{\gamma'}^{2} \,\tilde{X}_{\mu } \tilde{X}^{\mu }
+J^{\mu }A_{\mu }
\,,
\label{Eq.9}
\end{equation}
where $\chi$ is the dimensionless kinetic mixing strength, $m_{\gamma'}$ the DP mass and $J^{\mu }$ the EM current density. 

Dark photon is an interesting candidate for DM in the range $\chi \lesssim 10^{-11}-10^{-10}$. The exclusion regions for DP are shown in Fig. \ref{fig_8}. Details on the factors $\tau_2$, $N_{\nu}^{eff}$, CMB Distortions and others can be consulted in \cite{Arias_2012}.\newline

In Fig. \ref{fig_5} it is remarkable that the region in the QCD axion sector $10^{-5}<m_a<10^{-1}$ [eV] 
remains poorly explored. In Fig. \ref{fig_8} is shown an unexplored window for the detection of DPs. This experiment presents a potential to explore these sectors.

\section{Experiment set-up}
\label{sec:experiment}
DALI was thought to operate in three different modes: \textit{haloscope mode}, in which the instrument is stationary, 
and which focuses on the detection of relic axions forming part of the galactic halo; \textit{tracking mode}, which benefits from the adding of an altazimuth platform that gives the 
detector the capacity of pointing and tracking sources on the sky; and \textit{raster mode}, in which regions of 
the sky are covered with an algorithm that permits the scanning of wide areas searching 
for sources. Tracking and raster modes are \textit{telescope mode}.

\subsection{Theoretical foundation}
\begin{figure}[ht]
		\includegraphics[width=0.55\textwidth]{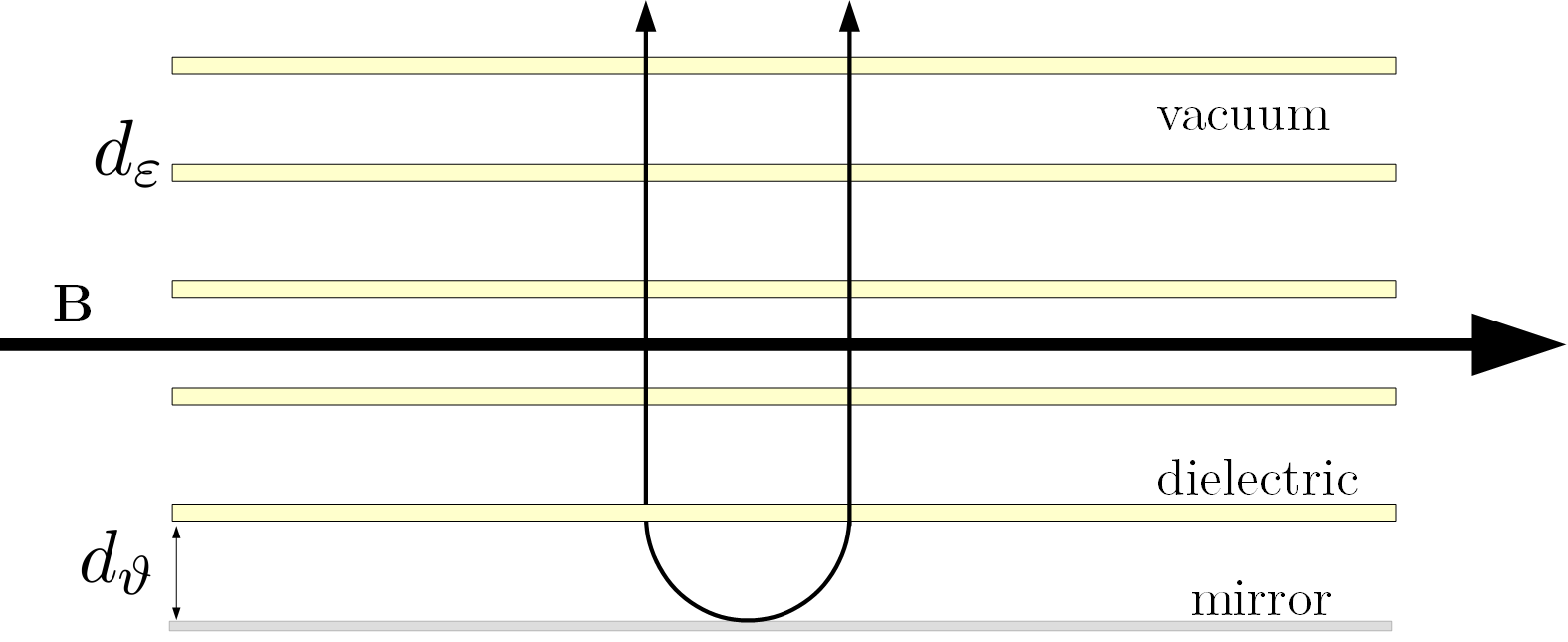}
		\centering 
		\caption{Experiment schematic. Axion-induced semi-plane waves are generated in the interface between vacuum and dielectric. The mirror at the bottom and a smart distribution of the plates can be used to create constructive interference. Consecutive plates act as a resonator producing power enhancement in a narrow band. The output can be received using microwave radiometers. The thickness between consecutive plates or between the bottom plate and the mirror
 ($d_{\vartheta}$) and the thickness of the dielectric plates ($d_{\varepsilon}$) must be
 adjusted with precision to maintain phase coherence.}
		\label{fig_0}
	\end{figure}

The specific case of a dielectric interface surrounded by a vacuum within an external magnetic field shown in Fig. \ref{fig_0} is 
treated now. A modification of Maxwell's equations can be derived arising from a light, pseudo-stable QCD 
axion \cite{PhysRevLett.51.1415}. From a classical approach, the axion mix with photons within an EM field with a Lagrangian density

\begin{equation}
\pazocal{L}_{a\gamma}=\mathord{\mathrm{g}}_{a\gamma} \, a \, \mathord{\mathrm{E}} \cdot \mathord{\mathrm{B}}  \;,
\label{Eq.1}
\end{equation}
where E and B are the electric and magnetic field, respectively. Within the classic limit, the axion field 
can be approximated by $a=\theta_0 \mathord{\mathrm{cos}}(m_at) \, f_a$, where \textit{t} is time and $\theta_0\simeq 4\times10^{-19}$ \cite{Millar_2017}.

Owing to the action of the magnetic field, a density current ($J$) enters on the right-hand side of Ampere's law in the interface between the vacuum and dielectric

\begin{equation}
J= - \mathord{\mathrm{g}}_{a\gamma} \, f_a \, \mathord{\mathrm{B}} \, \mathord{\mathrm{\dot{\theta}}}   \;,
\label{Eq.2}
\end{equation}
where $\mathord{\mathrm{\theta}}=\theta_0 \mathord{\mathrm{cos}}(m_at)$. This generates an electric field (E) in the interface

\begin{equation}
\mathord{\mathrm{E}}=\int J \cdot ds=\frac{f_a}{\varepsilon} \, \mathord{\mathrm{g}}_{a\gamma} \, \mathord{\mathrm{B}} \, \mathord{\mathrm{\theta}} (t)   \;.
\label{Eq.3}
\end{equation}

 The application of continuity conditions to parallel boundaries in the interface between both media in
 Eq. \ref{Eq.3} ($\mathord{\mathrm{E_1}}=\mathord{\mathrm{E_2}}$ \& $\mathord{\mathrm{B_1}}=\mathord{\mathrm{B_2}}$)
 shows that EM waves are generated to compensate for the discontinuity in electric permittivity
($\varepsilon_1 \neq \varepsilon_2$). Near the \textit{zero velocity limit} of axions, the momentum parallel 
to the interface is conserved, whereas the perpendicular momentum is determined by the dispersion relations, forcing 
the EM waves to be emitted perpendicularly to the interface \cite{Millar_2017_2} with a frequency $\nu_a=m_a/2\pi$ 
\cite{Horns_2013}.\footnote{Natural units are used throughout the text.} A quantum-field theoretical derivation has been treated more recently for the same case of a planar dielectric
 interface surrounded by a vacuum, obtaining a similar result to that the used throughout this work \cite{Ioannisian_2017}.

The relation between $f_a$ and $m_a$ is (e.g., see \cite{Millar_2017})

\begin{equation}
m_a=5.70(6)(4) \, \mathord{\mathrm{\mu eV}} \frac{10^{12} \, \mathord{\mathrm{GeV}}}{f_a}
\,,
\label{Eq.6}
\end{equation}
where the numbers in brackets account for the uncertainty. 

The wavelength of EM radiation emitted from the surface is given by the axion oscillation pulse $\omega\sim m_a$. Thus

\begin{equation}
\lambda_a\simeq\frac{2\pi}{m_a}\, (\mathord{\mathrm{eV}})({1.97\times10^{-7}\,\mathord{\mathrm{m}}})
\,. 
\label{Eq.7}
\end{equation}
\setlength{\belowdisplayskip}{1pt}\newline

The calculation of the wavelength and mass relation of DPs is analogous to Eq. \ref{Eq.7}. 

\subsection{Experimental approach}

Several apparatus for axion and DP detection have been suggested. We highlight the dish-antenna (\cite{Horns_2013}, 
\cite{Dish_antenna}), cavity resonator haloscope \cite{PhysRevLett.51.1415} and Fabry--P\'erot (FP) haloscope, including the so-called dielectric haloscope concept, adopted by the DALI experiment (\cite{Orpheus},
 \cite{PhysRevLett.118.091801}, \cite{Brun2019}).\footnote{Cavity resonator haloscopes are ongoing experiments.} 
Experiments based on LC resonant circuits also present a remarkable potential for the search for ALPs, although they are generally shorter in 
bandwidth, limited to low frequencies and relatively weak in sensitivity.\footnote{See \cite{Melc_n_2018} for a 
state-of-the-art proposal facing these limitations.}

We have established the exclusion sectors for axion, ALPs and DP detection in Figs. \ref{fig_5} and \ref{fig_8}. These graphics shown significant under-explored regions in both mass range or frequency and coupling strength. The origins of haloscopes are based on the need for power 
enhancement of the signal generated by axions and ALPs mixing with photons \cite{PhysRevLett.51.1415}. That is, 
the EM output produced by the physical mechanisms explained before in this manuscript is too weak owing to the 
limited magnetic inductance in real experiments (or low mixing strength, density) and must be enhanced before making 
possible a direct detection in a reasonable integration time.\footnote{Of the order of $\mathord{\mathrm{10^{-27}}}$W/$\mathord{\mathrm{m^2}}$ 
considering axions forming the galactic halo in the Solar System position \cite{PhysRevLett.51.1415} to be 
increased to around $\mathord{\mathrm{10^{-23}}}$W/$\mathord{\mathrm{m^2}}$.}
In the pre-inflationary scenario, resonant cavity 
haloscopes are adequate. However, the  dimensions of a resonant cavity scale with the wavelength, and cavity resonator haloscopes lose DB coherence when $m_a > 40\,\mathord{\mathrm{\mu eV}}$ 
\cite{Irastorza_2012}, or they are too large in the radio-wave domain, where different set-ups are used 
(e.g., \cite{PhysRevLett.122.121802}). The post-inflationary mass range is accessible using FP haloscopes, and 
it is the aim of our experiment to explore this concept.\footnote{In the limit between pre- and 
post-inflationary scenarios, both resonant cavity and FP haloscopes can be used.} 

In general, the principle of an FP interferometer is based on an interference between incoming and reflected waves 
within a resonator, forming a standing wave. Constructive interference occurs when incident and reflected waves
 are in phase. The output is spectrally modified
 compared to the input beam, allowing power enhancement in relatively narrow frequency bands centered at a resonant frequency. The 
classic FP resonator consists of a pair of mirrors surrounded by vacuum, or a medium with a known refractive 
index ($n=\sqrt{\varepsilon_r}$). FP interferometers and etalons have been used frequently in astronomy and are
 a very well established technology for infrared observations, where the mechanical requirements are relaxed 
compared to the optical range \cite{FP_review}.

The classical theory on FP resonators has already been applied to the case of a dielectric (FP) haloscope 
(\cite{Millar_2017}, \cite{MADMAX_2017}), where the \textit{internal resonance enhancement factor} 
of the classic FP interferometer or the quality factor ($Q$) play a role in the so-called \textit{boost factor} ($\beta$).\footnote{They differ by a factor that scales with the number of plates and its electric permittivity.} Throughout this work we follow this
 nomenclature. The boost factor is a figure of merit expressing the signal enhancement referred to a single 
magnetized mirror.\footnote{Of area equal to the area of a plate, or equivalently the area of the mirror.} The 
power boost factor scales with the number of stacked plates (N) and its electric permittivity, 
$\beta^2(\nu, \mathord{\mathrm{N}}, \varepsilon_r)$. In the proof-of-concept experiments, a narrow fake axion signal is injected 
in a continuum, and then detected with $\sim$5$\sigma$ significance using high-electron-mobility transistor 
(HEMT) radiometry (\cite{Brun2019}, \cite{proof_2016}, \cite{egge2020proof}).

Since DPs do not need the action of an external magnetic field to mix with SM photons, the observation of DPs maintains polarization coherence. In contrast, polarization in axion searching is degraded 
by the effects of the external magnetic field in haloscopes. \newline

\begin{figure}[ht!]
\centering
  \begin{minipage}{0.48\textwidth}
    \centering
    \includegraphics[width=0.7\textwidth]{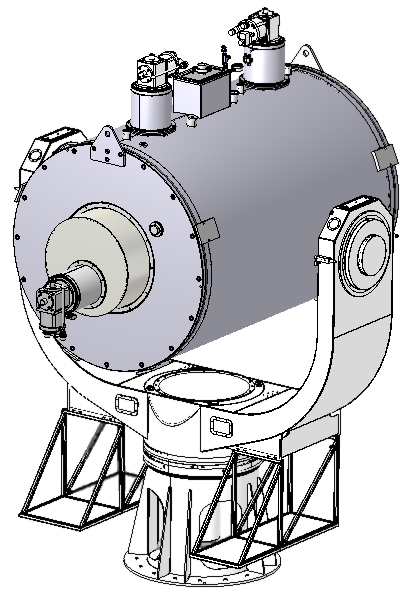}
     \put(-160,120){1}
    \put(-155,123){\vector(10,6){20}}
    \put(-115,200){2}
    \put(-109,202){\vector(10,-1){30}}
   
    \caption{Experiment set-up. General view. DALI consist on a Fabry--P\'erot interferometer within a cryostat. 
This cryostat is housed within a multicoil superconducting magnet ("1"). Approximate dimensions are 1.5x$\phi$1 m for the magnet 
and 2x$\phi$0.5 m of the experiment cryostat ("3"). }
    \label{fig_1}
  \end{minipage}%
  \hspace{5mm}
  \begin{minipage}{0.48\textwidth}
    \centering
    \includegraphics[width=0.995\textwidth]{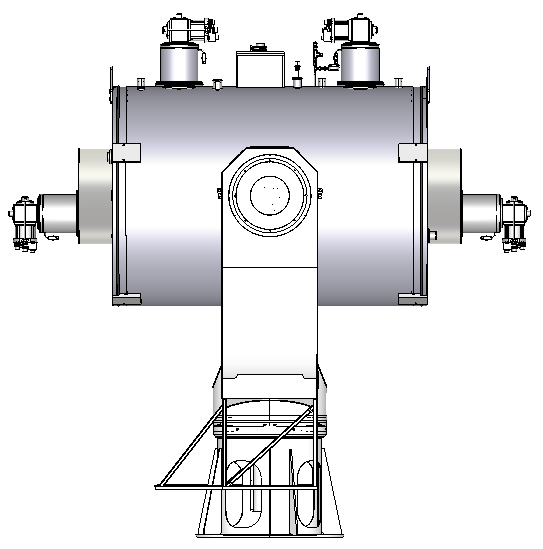}
     \put(-195,58){3}
    \put(-190,66){\vector(10,23){20}}
     \put(-32,15){7}
    \put(-33,19){\vector(-10,8){40}}
    \caption{Experiment set-up. Lateral view. Symmetrical design allows one to improve the thermal stability for the experiment cryostat ("3").  The apparatus rests on an altazimultal mount ("7"), which 
permits pointing and tracking with high speed rotation. Some objects has been removed for simplicity.}
    \label{fig_3}
  \end{minipage}
\end{figure}

\begin{figure}[ht]
    \centering
    \includegraphics[width=0.6\textwidth]{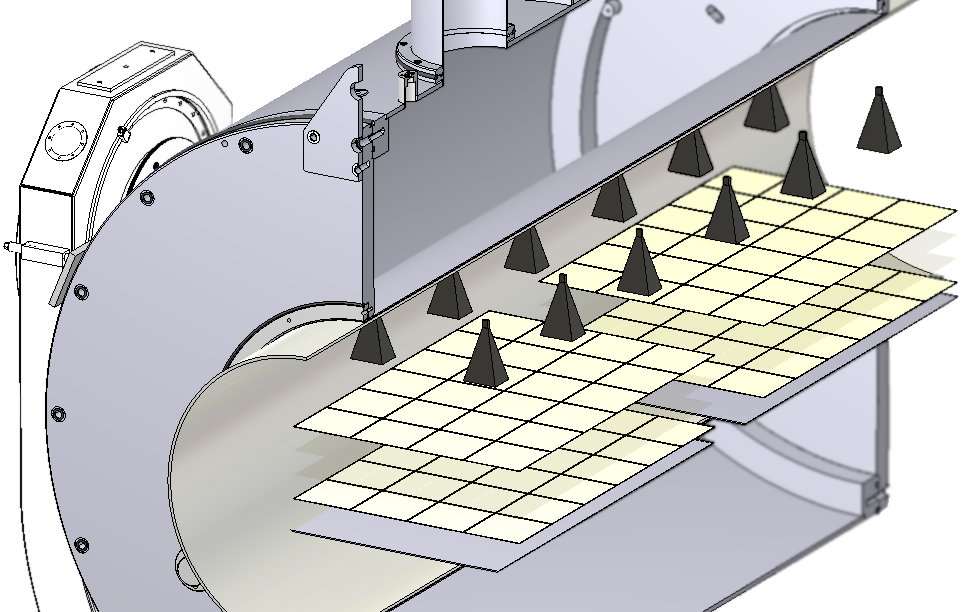}
    \put(-200,40){4}
    \put(-194,44){\vector(1,-0.11){10}}
    \put(-200,20){5}
    \put(-194,24){\vector(1,-0.09){10}}
     \put(-10,145){6}
    \put(-10,145){\vector(-1,-1.1){8}}
    
\caption[Fig2]{Experiment set-up. Section. Detail of resonator ("4" and "5") and antennas ("6"). The grid ("4") is resizable so that it can  
be adjusted to the de Broglie (DB) wavelength of the axion, acquiring coherence. Here a grid size for
 maximal sensitivity to virialized DM particles and accelerated particles up to velocities of the order of a few
 percent speed of light is shown.}
\label{fig_2}
\end{figure}
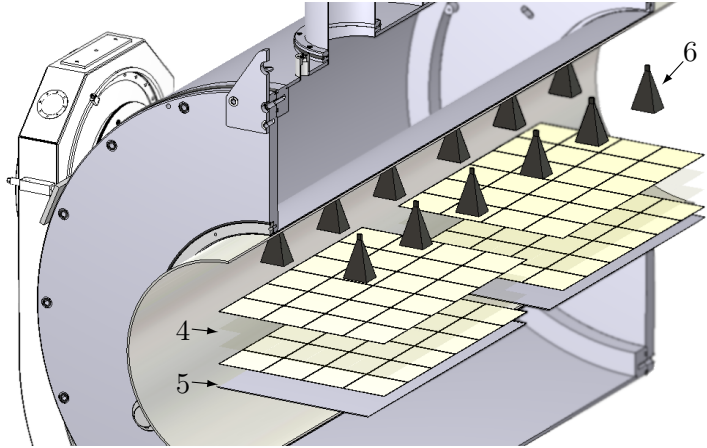

The DALI set-up is shown in Figs. \ref{fig_1} to \ref{fig_2}. A multicoil superconducting magnet ("1") houses 
the FP interferometer and microwave receivers. Multicoil magnets are commonly used in magnetic resonance imaging (MRI), 
industry and research. They reach typical magnetic inductances of 3, 5, 7 and 9 T, and eventually higher. The bore size 
is typically around 0.5 m with magnetic-field lines parallel to the cylinder axis. In a regular model, the length of 
the coil winding part is around 1.5 m, while the magnet total length is about 2 m. Field homogeneity is typically 1\% 
over a sphere of 100 mm diameter within the magnet bore and field stability is around 3 ppm/h. The superconducting 
magnet uses an independent helium free cooling system ("2"). The experiment cryostat ("3") is contained within the magnet's bore, but is independent. The cryostat is fabricated 
in non-magnetic material (Al).\footnote{Ti and fiberglass are also non-magnetic materials commonly used in
 cryogenics.} In order to achieve high thermal stability and homogeneity, the cryostat is mounted symmetrically 
and equipped with twin cold-heads. 
The primary mirror is composed of a group of stacked dielectric parallel plates ("4"). These plates form the
 FP interferometer. A polished mirror is attached at the bottom ("5"). The dielectric plates are composed 
of a grid made of the union of commercial wafers. A higher electrical permittivity results in a higher signal power.\footnote{We suggest to use zirconia ($ZrO_2$), a commercial 
material with dielectric constant $\varepsilon_r\simeq29$ and loss tangent $\mathord{\mathrm{tan}}\,\delta\sim10^{-4}$.} Standard dielectric wafers dimensions are 100 x 100 
mm size of 0.25, 0.5, 0.75, 1 and 1.25 mm thicknesses. The polished surface roughness is around 0.02 $\mu$m. Each plate is mounted on a holder. Stacking wafers allows one to benefit from combining and intermediate thicknesses. 
This allows us to set a different plate thickness in every sub-band (or group of sub-bands) over the 6--60 GHz band,
keeping $\beta^2$ not far from ideality. In order to benefit from maximum boost power, two thicknesses must be
 adjusted with precision: the plate spacing, which represents the thickness between consecutive plates or between the bottom plate and the mirror
 ($d_{\vartheta}$); and the thickness of the dielectric plates ($d_{\varepsilon}$). Both $d_{\vartheta}$ and
 $d_{\varepsilon}$ scale with wavelength. The theoretical thickness of the dielectric plate $d_{\varepsilon}$ varies from around 2.5 mm at 6 GHz (or $m_a\sim$25 $\mu$eV) to 0.25 mm at 60 GHz (or $m_a\sim$250 $\mu$eV) when working near transparent mode.
 The vacuum thickness $d_{\vartheta}$ varies from around 25 mm at 6 GHz to 2 mm around 60 GHz. These have a 
remarkable consequence: since the dimensions of the magnet's bore are fixed, more dielectric plates can be 
stacked at higher frequencies, increasing the boost power. Thus, the instrument should be
 more sensitive at higher frequencies in the band, although there the quantum noise is also higher, compromising the benefit in sensitivity. To achieve a high boost factor, the sub-bands must be narrower at lower frequencies. The $\beta^2$ feature is narrow band because 
$P\Delta \nu_{\beta}$ is roughly constant. The plan is to scan in sub-bands of the order of $\Delta\nu_{\beta}\sim100$ MHz wide, 
where $\beta^2\sim 10^4$ is achievable in practice \cite{PhysRevLett.118.091801}. Since the sub-bands are determined before starting observations, 
a reliable optimal plate spacing can be calculated in advance using finite-element-method (FEM) 3D simulation. 

For plate spacing, electro-mechanical actuators are used. Non-magnetic piezoelectric motors
have been already proved in axion magnetized experiments (\cite{PhysRevLett.121.261302}, \cite{zhong2017recent}). 
They present an uncertainty in the positioning of the order of $\pm$2 $\mu$m. This uncertainty is in tension with the 
experiment's tolerance, of the order of a few $\mu$m at 60 GHz (and tens of $\mu$m at 6 GHz).\footnote{For 90\% 
efficiency.} Static deflections and other systematic errors must be considered as well. However, part of the loss due to mechanical deviation would be absorbed by FP reconfiguration 
based on 3D--FEM EM simulations, in principle. The same logic is applicable to edge spurious effects and permittivity
 discontinuities in the dielectric grid, based on the results in \cite{schtteengel2018simulation} and \cite{Knirck_2019}.

Receivers can cover the focal plane of a telescope through different strategies. A classic schema incorporates a focusing mirror which concentrates all the photons in a reduced focal plane, and places a single detector that covers it completely. This is the case of reflector antennas (\cite{Brun2019}, \cite{Miguel_Hern_ndez_2019}). Differently, phased arrays \cite{Mailloux} strategically place a number of antennas that cover a fraction of the focal plane each in order to combine their outputs. This is the case of focal plane arrays. This set-up has been used by numerous radio-telescopes over decades (\cite{Watson:2002tf}, \cite{Gregorio:2013uya}, \cite{Quijote}, \cite{Lee:2020usv}, \cite{Incardona:2020nuo}). Both strategies are equivalent, ideally. Since sacrificing space within the magnet bore to incorporate a parabolic reflector would reduce the room available for plate stacking, DALI adopts the second configuration at the cost of introducing heavier electronics. Thus, the output power is received by an antenna array placed at the top of the cryostat ("6"). The signal is amplified 
in independent pixels using HEMTs\footnote{We discuss an alternative technology in section \ref{sect:sensitivity}.} 
and processed in a cold front-end module. The signal is then  transferred to a room-temperature back-end module, 
not shown in the schema for simplicity. The data acquisition system (DAS) is based on field-programmable gate 
array (FPGA) structure. The signal maintains phase coherence and the radiometer uses a pseudo cross-correlation schema. Cross-correlation allows one to mitigate the gain fluctuations (or equivalently noise temperature) by comparison between two uncorrelated signals \cite{Faris}. A similar effect can be obtained by combining the signal from multiple outputs in a multi-receiver system. Hence, 
the mitigation of low frequency gain fluctuations caused by thermal and 1/\textit{f} noise contributions is
 significant. Receivers could be pointed to cold-loads as a reference calibration signal, if necessary (e.g., see \cite{De_Miguel_Hernandez_2019} and references therein). This might render the directional observation of relativistic axions possible, even in specific cases where the line width of the axion\footnote{$\Delta\nu_a/\nu=\sigma_{\mathrm{v}_a}^2/2$.} (or DP) may be considerably wide, approaching $ \Delta \nu_a / \nu \sim 1$ in extreme cases. 

An altazimuth mount ("7") equipped with a rotating joint \cite{2012SPIE.8452E..1MT} transmits sufficient torque for 
fast rotation of the approximately 3 mT weight of the instrument, making pointing and tracking feasible.\footnote{The
 design shown in Fig. \ref{fig_1} is adapted from \cite{Quijote} and could undergo modifications.} 

The experiment's laboratory must be isolated from spurious microwave backgrounds. A standard Faraday cage
 provides around 100--120 dB attenuation over 10 MHz and up to frequencies of tens of
 GHz.\footnote{\href{url}{http://www.hollandshielding.com}} \newline

Since DM is weakly interactive, the telescope is bidirectional and some uncertainty in the pointing model is
 present. However, FP haloscopes are theoretically sensitive to the side of the plate in which the axion
 momentum is transferred trough phase information \cite{Knirck_2018}. The line of sight (l.o.s.) of the 
instrument, perpendicular to the dielectric plates, bifurcates in l.o.s$^+$ and l.o.s$^-$ passing through the 
Earth. This allows simultaneous observation in both hemispheres.

\subsection{Sensitivity projection}
\label{sect:sensitivity}

\begin{figure}[t!]
		\includegraphics[width=0.65\textwidth]{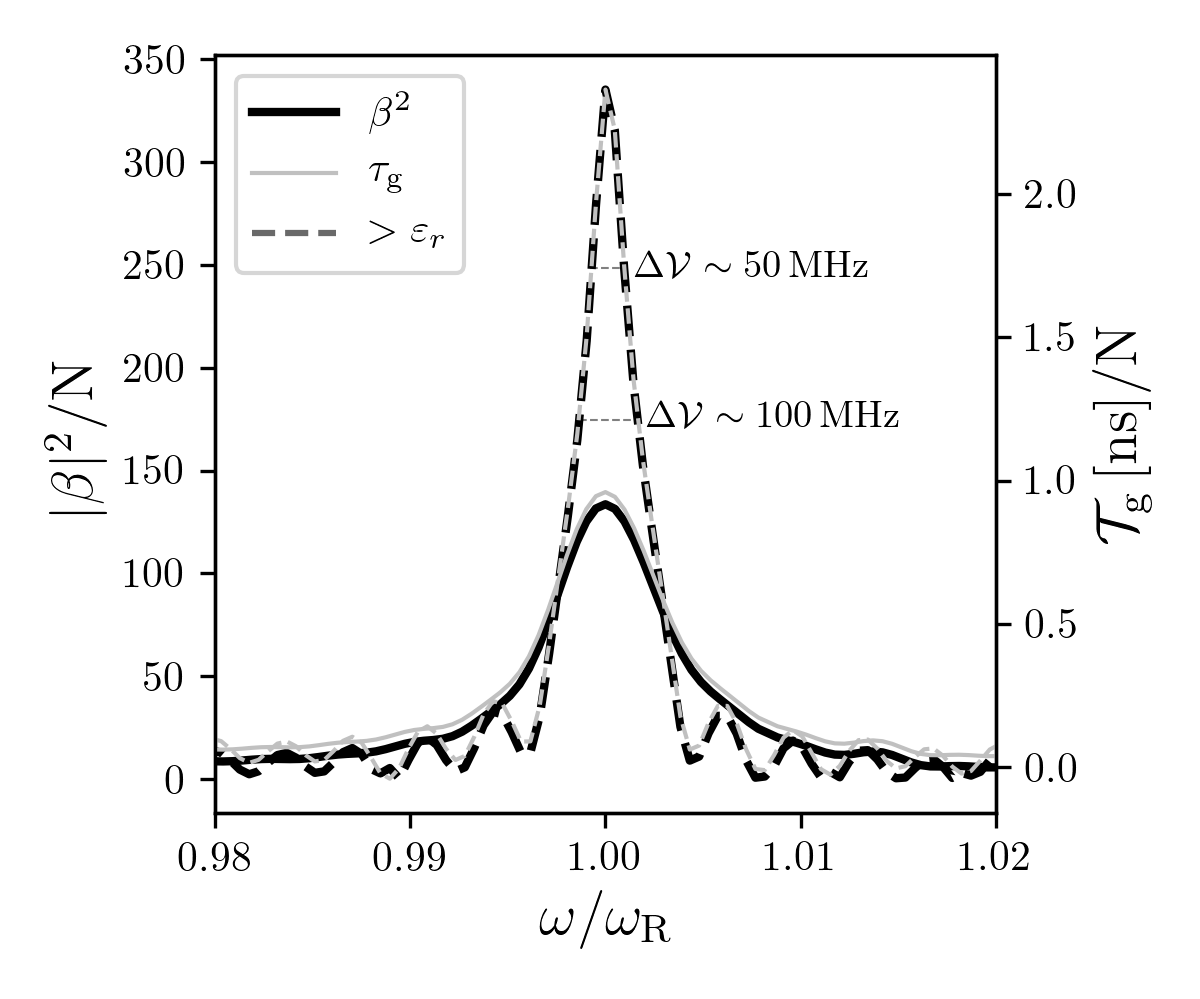}
		\centering 
		\caption{Power boost factor ($\beta^2$, in black) as a function of group delay time ($\mathord{\mathrm{\uptau_g}}$, in gray) around the resonant (pulse or equivalently) frequency ($\omega_R$) normalized to the number of stacked plates (N). Results from 3D-FEM simulations for a parsimonious model where $\varepsilon_r$= 9.4 ($Al_2O_3$, solid line) or 29 ($ZrO_2$, dashed line), $d_{\varepsilon}$=1 or 1.5 respectively, and $d_{\vartheta}\sim$ 7 mm. A higher value of $\varepsilon_r$ results in a higher peak for $\beta^2$ ($\propto \varepsilon_r$) and a narrower full width at half maximum (FHWM) ($\propto \varepsilon_r^{-1}$). Two representative band widths ($\Delta \nu$) are highlighted as a reference for the estimate of the sensitivity of the experiment. The model is consistent with the experimental results presented in \cite{Brun2019}.}
		\label{fig_6}
	\end{figure}

\begin{figure}[t!]
		\includegraphics[width=0.7\textwidth]{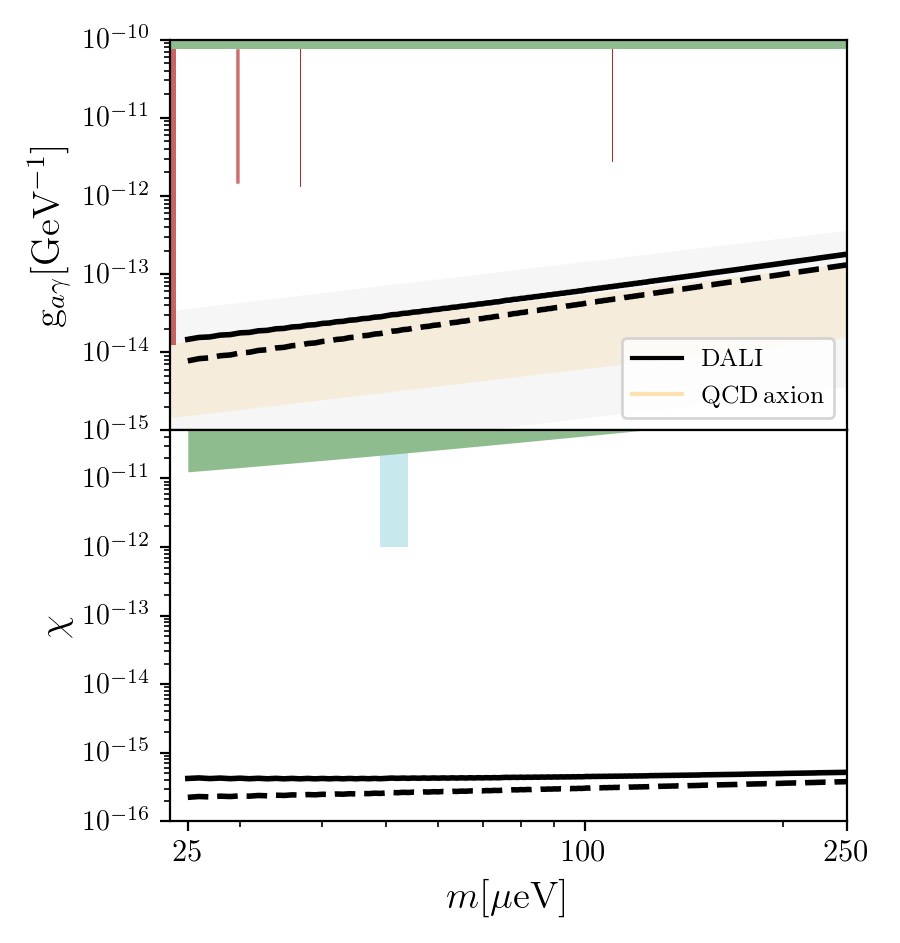}
		\centering 
		\caption{Sensitivity projection to relic axions (upper) and dark photons (lower) in the 
6--60 GHz band. Parameters are B$\sim$9 T, data are stacked over a few-month period by adding shortened integration times in order to make the experiment sensitive to the background modulation, $\rho_{DM}\sim0.45$ GeVcm$^{-3}$, $\alpha$ corresponds to random velocity dispersion, 3$\sigma$ significance 
and $T_{sys}$ present two different cases: $^3$He--$^4$He 
sorption cooler down to a physical temperature about 1 K and HEMTs with a noise temperature about 2--3 times the quantum limit (solid line); and a bath temperature of 50 mK with ultra-low noise quantum detectors with a characteristic noise temperature about 1.5 times the quantum limit (dashed line). Sensitivities are compared to QCD axion models over the entire 25--250 $\mu$eV range. Exclusions regions in the vicinity are represented in green for horizontal branch (HB) and haloscopes in red (\cite{PhysRevD.97.092001}, \cite{MCALLISTER201767}, \cite{Alesini:2019ajt}). The Tokyo limit is shown in blue \cite{suzuki2015hidden}. The axion-model window in light gray is defined in \cite{DiLuzio:2016sbl}.}
		\label{fig_9}
	\end{figure}

The sensitivity projection of DALI equipped with a 9 T multicoil superconducting magnet is shown in Fig. \ref{fig_9} for the specific configuration of interferometer presented in Fig. \ref{fig_2}. In Fig. \ref{fig_9} we consider two different cases: (i) double stage $^3$He--$^4$He sorption cooler working with
a background physical temperature of 1 K for HEMT based detectors \cite{McCulloch} and a noise temperature consistent with the heat dissipation limit established by \cite{Schleeh}, or (ii) sub-kelvin operation replacing HEMTs with
 ultra-low-noise quantum devices (\cite{2011APS..CAL.F2007C},
 \cite{ASZTALOS201139}, \cite{2016APS..APRK16004O}, \cite{8990953}, \cite{MATLASHOV2018125}).\footnote{The technology for ultra-low-noise detection involving high-frequencies is beyond the state-of-the-art.} The number of plates is given by N$\sim$ L$/(d_{\varepsilon}+d_{\vartheta}$), where L is the length available for plate stacking inside the magnet bore, around 300 mm.\footnote{As $d_{\vartheta}\sim \lambda/2$, the most unfavorable frequency is 6 GHz, where 12 plates are stacked when working in transparent mode. A study is underway to determine if the plate spacing can be shortened without compromising the boost factor.} A study of the power boost factor incorporated to the estimate presented in Fig. \ref{fig_9} is included in Fig. \ref{fig_6}, where 3D-FEM simulations are presented for a parsimonious\footnote{Simplified, scaled, etc., whose results are transferable to the case of study.} model of the interferometer. The group delay ($\mathord{\mathrm{\uptau_g}}$) informs on the mean lifetime of the photons within the interferometer in the resonant mode \cite{Renk}. The quality factor of each individual cavity forming the interferometer scales as $Q=\omega \times  \mathord{\mathrm{\uptau_g}}$ and the power boost factor around the resonant frequency is $\beta^2\propto \, Q$. The boost factor scales linearly with the number of stacked plates ($\beta^2\propto Q\propto\,$N) \cite{Millar_2017}. Consequently, the boost varies between $\beta^2\sim10^{3-4}$ within the experimental range. Systematics are studied in \cite{Beurthey:2020yuq}, \cite{proof_2016} and \cite{egge2020proof}. \newline 
 
The sensitivity projection to relic axions and ALPs of DALI operating in haloscope mode over the entire 25 to 250 $\mu$eV 
range is shown in Figure \ref{fig_9} (upper). The sensitivities are compared to theoretical predictions of reference axion models. In Eqs. \ref{Eq.12} to \ref{Eq.14} $A$ is the collecting area, $\mathord{\mathrm{B}}$ the external 
field, $k_B$ the Boltzmann constant, $T_{sys}$ the system temperature, $\Delta \nu \equiv \Delta \nu_a=10^{-6}\nu$ the 
bandwidth referred to axion line width and $t$ is integration time. The axion--photon coupling factor is given by (e.g., see \cite{Millar_2017})

\begin{equation}
{\mathord{\mathrm{g}}}_{a \gamma} = 2.04(3)\times10^{-16} \, C_{a \gamma} \, \frac{m_a}{\mu{\mathord{\mathrm{eV}}}} 
\,.
\label{Eq.12}
\end{equation}

The axion-induced electric field generates an energy flux density (\cite{Millar_2017}, \cite{MADMAX_2017}, \cite{proof_2016}) 

\begin{equation}
\frac{P_{a\gamma}}{A}=2.2\times10^{-27}\,\beta^2\,\left(\frac{\mathord{\mathrm{B}}}{10\,\mathord{\mathrm{T}}}\right)^2\,C^2_{a\gamma} \,\left(\frac{\rho_{DM}}{0.3\,\mathord{\mathrm{GeVcm^{-3}}}}\right)
\,,
\label{Eq.13}
\end{equation}
while the received power can be estimated using the well known \textit{ideal radiometer equation} \cite{Dicke} in the form

\begin{equation}
\frac{S}{N}=\frac{P}{k_B\,T}\sqrt{\frac{t}{\Delta{\nu}}}
\,.
\label{Eq.14}
\end{equation}
\setlength{\belowdisplayskip}{0pt}

The sensitivity to virialized DPs of the experiment over the 25--250 $\mu$eV band is shown in Fig. \ref{fig_9} (lower). In Eq. \ref{Eq.10} $\rho_{DM}$ is DM halo density, $A$ the detection area and $\alpha$ 
is a factor determining the incidence angle of the DP ($\alpha=\sqrt{2/3}$ represents random case) \cite{Horns_2013}. \textit{P} 
can be obtained from Eq. \ref{Eq.14} being $\Delta \nu \equiv \Delta \nu_{DP}=10^{-6}\nu$.

\begin{equation}
\mathord{\mathrm{\chi=4.5\times 10^{-14}\left( \frac{\textit{P}}{10^{-23}\,W} \frac{0.3 \,GeVcm^{-3}}{\rho_{DM}} 
\frac{m^2}{A} \frac{1}{\beta^2} \right) ^{1/2} \frac{\sqrt{2/3}}{\alpha}}} \,.
\label{Eq.10}
\end{equation} \newline

The experiment shows a potential to probe unexplored axion (and ALPs) and DP sectors over a broad band of frequencies.

\section{The aim for directional observation}
\label{sec:Directional}
The mathematical formalism and first approximations to establish a calculation of sensitivities for the case of 
low-velocity axions in stationary directional haloscopes have been elegantly carried out already in (\cite{Millar_2017_2}, 
\cite{Knirck_2018}). Ongoing stationary low-frequency experiments incorporated some of these uncertainties some time ago, 
anticipating the possibility of receiving microwave photons from random axion events (\cite{Duffy:2006via}, 
\cite{Duffy_2009}, \cite{Choi:2005rn}, \cite{PhysRevD.94.082001}). Helioscopes can be reconfigured in order to be 
sensitive to streaming CDM axions and miniclusters \cite{2017arXiv170301436Z}. On the other hand, the velocity distribution of the galactic DM halo cannot be ideally isotropic, so directionality 
could be relevant even at this basic level.  \newline 
The potential of this experiment to explore several theories and hypothesis involving astrophysical axion sources or the existence of DM substructures, such as tidal-streams and mini-clusters, in a 100\% compatible and simultaneous mode to the conventional search for virialized axions forming the galactic-halo performed by classic haloscopes, relies on the aspects treated throughout this section. 

\subsection{Detectability of non-virialized particles}
The velocity 
dispersion from the zero velocity limit up to the virial velocity around $10^{-3}$ through its 
effects on $\beta^2$ has been studied in a similar haloscope concept \cite{Millar_2017_2}. The 
detector becomes sensitive to velocity dispersion when the haloscope size is of the order of 
15--20\% of the axion DB wavelength. Below this limit, $\beta^2$ can  in principle be considered velocity 
independent and so the instrument presents low or null sensitivity to velocity dispersion and incoming 
direction of the axion. When the DB wavelength of the axion is of the order of the size of the haloscope,
 directional sensitivity is $\pazocal{O}(1)$. This gives a strong directionality to the \textit{axioscope}.
 Virialized axion search is weakly sensitive to directionality up to around 250 $\mu$eV.\newline 
 
The case of faster particles requires further analysis. The DB coherence must be maintained in order to make the axion detectable 
because of \textit{cross-section} considerations.\footnote{See \cite{Larkoski} pp. 86--87 for an 
intuitive review of this fundamental issue.} Therefore, the DB wavelength of the axion or ALP ($\lambda_{DB}=2\pi\gamma/m_a \mathord{\mathrm{v}}_a$)\footnote{$\gamma =1/\sqrt{1-\mathord{\mathrm{v}}^2/c^2}$ is the Lorentz factor.}
 must be larger than the \textit{collector} or \textit{primary mirror} scale length ($\ell$). This allows the axion field to be measured including phase information. Given this, the detector 
is coherent when the following condition is fulfilled

\begin{equation}
\mathord{\mathrm{v}}_{a_{\ell}} \lesssim \frac{2\pi}{\gamma\,m_a \, \ell}(\mathord{\mathrm{eV}})(1.97\times10^{-7}\,\mathord{\mathrm{m)}}\, 
\,.
\label{Eq.8}
\end{equation}

\begin{figure}[ht]
		\includegraphics[width=0.6\textwidth]{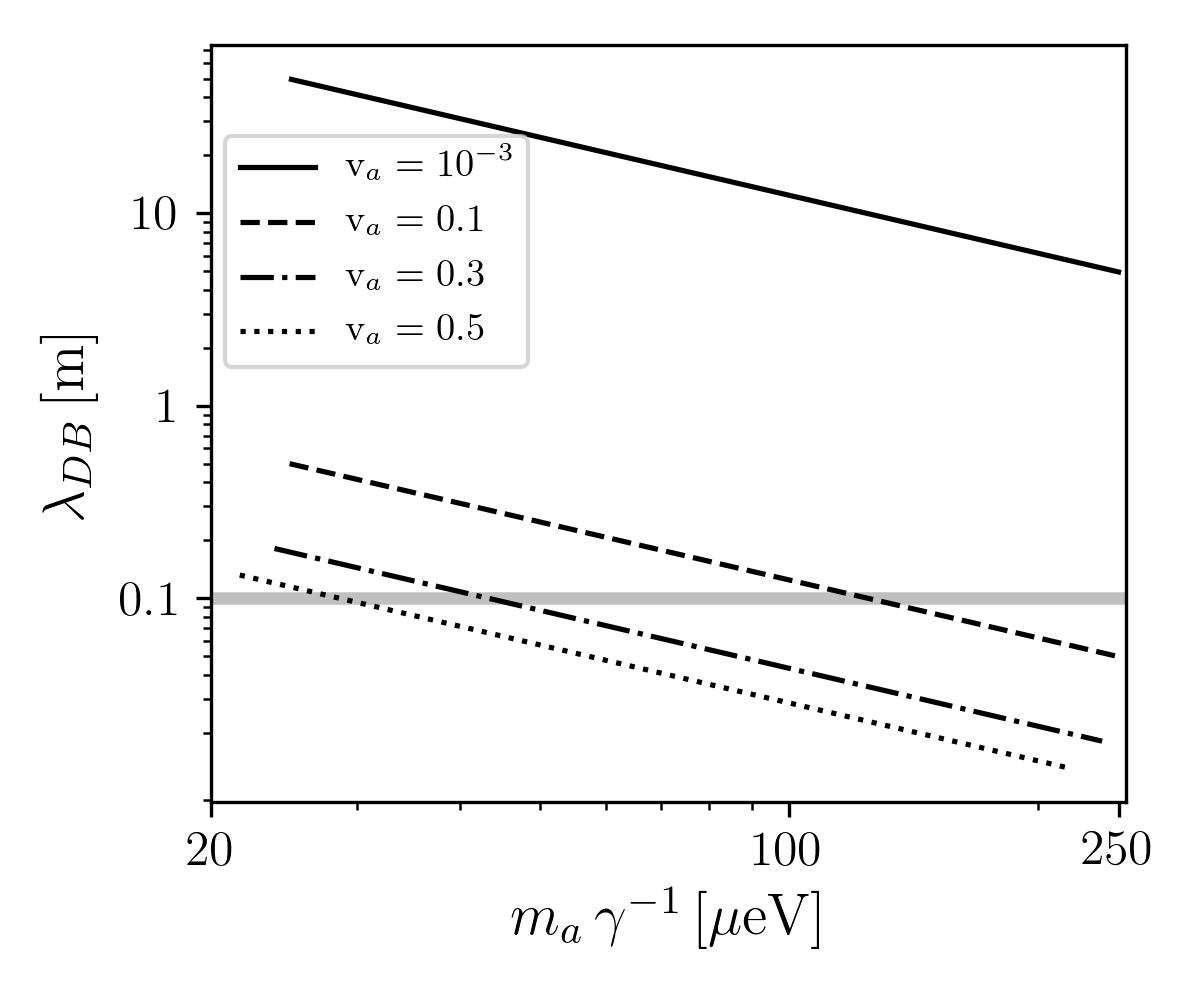}
		\centering 
		\caption{Study of de Broglie (DB) coherence in the frequency range 6--60 GHz. Limiting velocity ($\mathord{\mathrm{v}}_{a_{\ell}}$) over mass of detectable axions. The gray horizontal line represents the limiting DB wavelength, which is coherent with a basic unit with 
$\ell\sim0.1$ m. The region
 below the gray horizontal line loses DB coherence and is hidden. Axions with velocities up to around $0.5c$ are theoretically detectable with 
the more basic unit of grid of 10 cm length ($\ell=0.1$). The virial velocity of the relic axions, around $10^{-3}c$, as reference, and relativistic velocities up to $0.5c$, are represented.}
		\label{fig_7}
	\end{figure}
The expression in Eq. \ref{Eq.8} establishes a limit to the velocity 
of non-virialized axions that is possible to receive with a telescope characterized by the scale length of 
its detector ($\mathord{\mathrm{v}}_{a_{\ell}}$). The limiting velocity scales inversely to axion mass. This can be examined with the help of Fig. \ref{fig_7}, where the horizontal gray line represents the experiment's limit in scale length, to be compared 
to the relativistic DB wavelength shown in ordinate axis. Note the sector above the 0.1 m characteristic length line, revealing that axion velocities up
 to around 50\% of the velocity of light in vacuo are accessible in the case of the basic unit
 of around 10 cm in length.\footnote{Several basic units that cover the entire longitudinal section of the 
bore can be placed if they are sufficiently separated in order not to interfere.} Lighter (i.e., $m_a<25\,\mu$eV) relativistic axions are shifted to the experiment band. However, at these energies, 
Lorentz relativistic correction to the axion rest mass below 15\% ($\gamma\lesssim1.15$) remain moderate, 
specially considering the limited spectral resolution of a FP haloscope. Many basic plate units can be stacked and paired to form bigger and thicker collector areas. 
Some examples are the cases of 2, 5 and 10 paired basic units (a basic unit presents $\ell\sim0.1$, as mentioned) with 
respective velocities of 25, 10 and 5\% speed of light at $m_a \sim 25 \, \mu$eV. The diffraction limit $\ell_d \sim \lambda_a$ does not conflict with the scale length throughout the experimental range, eluding significant diffraction on the resonator plates. Finally, a highly relativistic velocity would dramatically reduce the sensitivity of the experiment mainly due to its effect on the axion phase, affecting the coherence length \cite{Millar_2017_2}. A practical realisation designed to scan for relativistic axions would require a setup with a limited number of plates and so a lower power boost factor, forcing a larger integration time. This would limit the detectability of non-virialized particles to velocities of the order of a few tens percent the speed of light, except in specific cases of higher occupation number compared to the local DM density. \newline

\subsection{A telescopic search for axion}

Telescopic search for non-virialized DM particles does not compromise the classical (stationary) exploration for galactic halo axions and DPs. Both necessarily take place simultaneously during scanning. Particles of diverse origin could mix. An extra population of DM particles with similar \textit{dynamic} mass would only result 
in a signal enhancement and hence an enhanced significance of detection, so long as the DB coherence 
is maintained. This would allow the detector to access sensitivity sectors that were previously 
prohibited. A modulation of the background signal would also reveal the signature of DM. In this section we discuss several examples:

(i) the existence and detectability of DM substructures is an interesting hypothesis arising from anomalous large-scale observations and simulations. Axion streams density would vary between 0.3--30\% of local DM density ($\rho_{DM}$). Thus, they would be potentially detectable. Furthermore, streams would present a $10^6$ flux enhancement factor compared to the local DM density (i.e., $\rho_a\sim 10^6\rho_{DM}$) once aligned in the Sun-Earth l.o.s. focusing low velocity axions at the Earth's position \cite{NIST}. Moreover, streams may be enhanced by microlensing. Gravitational microlens caused by planets in the direction planet-Earth may provide $\rho_a\sim 10^6\rho_{DM}$, or around $\rho_a\sim 10^4\rho_{DM}$ in the case of the Moon \cite{Zioutas:2017klh}. Only once data are stacked during month-year periods planetary effects underlying streams would be revealed in broad band observations with shortened integration time. A different case of interest is the Galactic-center-Sun-Earth alignment, repeated annually. This effect would be enhanced every 8--9 years, when the Moon is aligned in the same l.o.s. Streaming DM could explain why both the terrestrial atmospheric ionisation and solar activity show dependence on the longitudinal position of the planets or lunar phase \cite{Bertolucci}. Referring now to small-scale substructures such as miniclusters, they could be disrupted forming tidal streams with flux density $\rho_a\sim 10\rho_{DM}$ \cite{Tinyakov:2015cgg}. Stream-crossing events may occur around 1 every 20 years with a few-day duration, so the probability
 of entering or leaving a DM minicluster during a measurement is small. However, miniclusters trapped by the Solar System during its formation \cite{Zioutas:2017klh}, with $\rho_a\sim 10^5\rho_{DM}$ and few-day events duration annually have been suggested. Such trapped substructures are within the observational bounds \cite{Frere:2007pi}.  Refer to \cite{Tinyakov_2016}, \cite{Dokuchaev2017}, \cite{Berezinsky_2013}, \cite {PhysRevD.93.123509},
 \cite{Tkachev2015}, \cite{PhysRevLett.119.021101}, \cite{PhysRevD.97.083502}, \cite{PhysRevLett.121.151301}, \cite{VISINELLI201864}, \cite{HOGAN1988228}, \cite{vogelsberger2010streams} and \cite{BERTOLUCCI201713} for a general view on the status of \textit{dark universe} research. 
The same concept 
on DM substructures and flows explained here for the case of axions and ALPs can be extended to the search for DPs, 

(ii) recently, the ANITA experiment, dedicated to measuring isolated impulsive radio signals originating from cosmic 
ray showers reflected in the Antarctica ice, reported the detection of two anomalous events, which are not 
explainable by cosmic rays (\cite{Gorham:2018ydl}, \cite{Aartsen:2020vir}). An explanation strictly within 
the SM conflicts with other experiments and theory. Different arguments over the years depend on further 
investigations and/or tend to be in tension with data (e.g. \cite{PhysRevLett.123.091102}, \cite{Shoemaker}, 
\cite{fox2018anita}, \cite{Huang:2018als}, \cite{PhysRevD.100.015031}). The possibility of instrumental 
artifacts must not be discarded either. However, authors have recently suggested that these events could be explained 
in terms of resonance in the ionosphere by the axion--photon conversion mechanism induced by axion \textit{flares} \cite{Esteban:2019hcm},

(iii) with reference to stellar physics, solar axions produced by the Primakoff effect are ultrarelativistic, emitted with keV-level
energies and so are invisible for our experiment \cite{Anastassopoulos:2017ftl}. However, 
axion quark nugget (AQN) theory \cite{Zhitnitsky_2003} suggests a mechanism for creating a significant population of 
axions in the solar corona released with velocities of 10--90\% of the speed of light \cite {PhysRevD.98.043013}. 
The mass relativistic corrections are moderate, so photons generated by AQN-induced axion mechanisms may be received in the 
microwave range. AQN-induced axions would mix with relic axions \cite{Liang:2018ecs}. AQN-induced axions present flux density $\rho_a\sim 10^{-5}\rho_{DM}$ and typical velocity around 0.5$c$. Linearity in velocity enhances the flux density of AQN-induced axions up to $\rho_a\sim 10^{-2}\rho_{DM}$ \cite{Graham:2013gfa}. Benefit from microlensing and perhaps also the halo modulation amplification time dependent factor would be helpful in finding these DM particles (\cite{VERGADOS201710}, \cite{Ge:2020cho}, \cite{Froborg:2020tdh}). The preferred mass for AQN-induced axions is $m_a\sim100$ $\mu$eV \cite{Budker:2019zka}. Low velocity gravitationally trapped axions produced by AQN mechanism have been also suggested \cite{Lawson:2019cvy}. \newline

It is remarkable that DM-halo axions, axion substructures, isolated impulsive axion-induced signals (i.e., flares) and AQN-induced axions and DPs might be simultaneously explored during observations. The detection of non-virialized particles is preferably directional. Note that from the dot product in Eq. \ref{Eq.1} it follows that the interaction is maximal when the axion incidence angle to a 
dielectric plane is perpendicular to the magnetic field, transmitting maximum momenta. Otherwise, the transmission is penalized by a factor $\mathord{\mathrm{cos}}\:\!\iota$, with $\iota$ the incidence angle, and the signal is reduced accordingly. The study of the signal modulation originated by random crossing events is an interesting case. Although the analysis in \cite{Knirck_2018} is devoted to a stationary directional search for axion, it can be extended to the case of a telescopic exploration for dark matter substructures. Dominated by celestial mechanics, the measured signal would be sinusoidally modulated. The daily signal modulation caused by an axionic event with a velocity $\mathord{\mathrm{V}}\!_a$ can be expressed in terms of the modulation parameters $c_0, c_1$ and $\phi$ in the form 

\begin{equation}
\mathord{\mathrm{cos}}\:\!\xi_a^{\pazocal{N,W,Z}} = \frac{\mathord{\mathrm{\hat{e}}}^\pazocal{N,W,Z} \cdot (\mathord{\mathrm{V\!_{lab}}}-\mathord{\mathrm{V}}\!_a)}{\mid\!\!\mathord{\mathrm{V\!_{lab}}}-\mathord{\mathrm{V}}\!_a\!\!\mid} \longleftarrow c_0+c_1\mathord{\mathrm{cos}}(\omega_d\,t+\phi)
\,,
\label{Eq.11}
\end{equation}
where  $\omega_d=2\pi/0.997$, $t$ is time expressed in days from January 1st and $\mathord{\mathrm{V_{lab}}}$ stands for the lab velocity in laboratory coordinates. For the case of stationary haloscopes featuring directional sensitivity, rectangular lab coordinates $\pazocal{N,W,Z}$ are adequate. The modulation parameters are

\begin{equation}
\overbrace{b_0\,\mathord{\mathrm{cos}}\lambda_{\mathord{\mathrm{lab}}}}^{c_0^{\pazocal{N}}}-\overbrace{b_1\, \mathord{\mathrm{sin}}\lambda_{\mathord{\mathrm{lab}}}}^{c_1^{\pazocal{N}}}\mathord{\mathrm{cos}}(\omega_d\,t+\overbrace{\phi_{\mathord{\mathrm{lab}}}+\psi}^{\;\:\phi^{\pazocal{N}}}\,)
\,,
\label{Eq.15}
\end{equation}

\begin{equation}
\aoverbrace[U]{b_1}^{{c_1^{\pazocal{W}}}}  \mathord{\mathrm{cos}}(\omega_d\,t+\overbrace{\phi_{\mathord{\mathrm{lab}}}+\psi-\pi}^{\;\:\phi^{\pazocal{W}}}\,)
\,,
\label{Eq.16}
\end{equation}

\begin{equation}
\overbrace{b_0\,\mathord{\mathrm{sin}}\lambda_{\mathord{\mathrm{lab}}}}^{c_0^{\pazocal{Z}}}+\overbrace{b_1 \,\mathord{\mathrm{cos}}\lambda_{\mathord{\mathrm{lab}}}}^{c_1^{\pazocal{Z}}}\mathord{\mathrm{cos}}(\omega_d\,t+\overbrace{\phi_{\mathord{\mathrm{lab}}}+\psi}^{\;\:\phi^{\pazocal{Z}}}\,)
\,,
\label{Eq.17}
\end{equation}
where an offset is cancelled while $b_0=\sigma_3\mathord{\mid\!\!\mathord{\mathrm{V\!_{lab}}}-\mathord{\mathrm{V}}\!_a\!\!\mid}^{-1}$, $b_1=(\sigma_1^2+\sigma_2^2)^{1/2}\mathord{\mid\!\!\mathord{\mathrm{V\!_{lab}}}-\mathord{\mathrm{V}}\!_a\!\!\mid}^{-1}$ and $\psi=\mathord{\mathrm{tan}}^{-1}(\sigma_1/\sigma_2)-0.721\omega_d-\pi/2$. The factors $\sigma_1=(-0.055,0.494,-0.868)\cdot \Upsilon$, $\sigma_2=(-0.873,-0.445,-0.198)\cdot \Upsilon$, $\sigma_3=(-0.484,0.747,0.456)\cdot \Upsilon$ arise in the conversion into lab coordinates; with $\Upsilon=\mathord{\mathrm{V_{\odot}}}+v_{\oplus}\,\mathord{\mathrm{cos}}(\tau_y)(0.994, 0.109, 0.003)+v_{\oplus}\,\mathord{\mathrm{sin}}(\tau_y)(-0.052, 0.494, -0.868)$ and $\tau _y=2 \pi \, (t-79)/ 365$. Note that a west pointing haloscope is not sensitive to $b_0$.\newline

\begin{figure}[t!] 
		\includegraphics[trim=11.2 10 5 40,clip=true, width=1.025\textwidth]{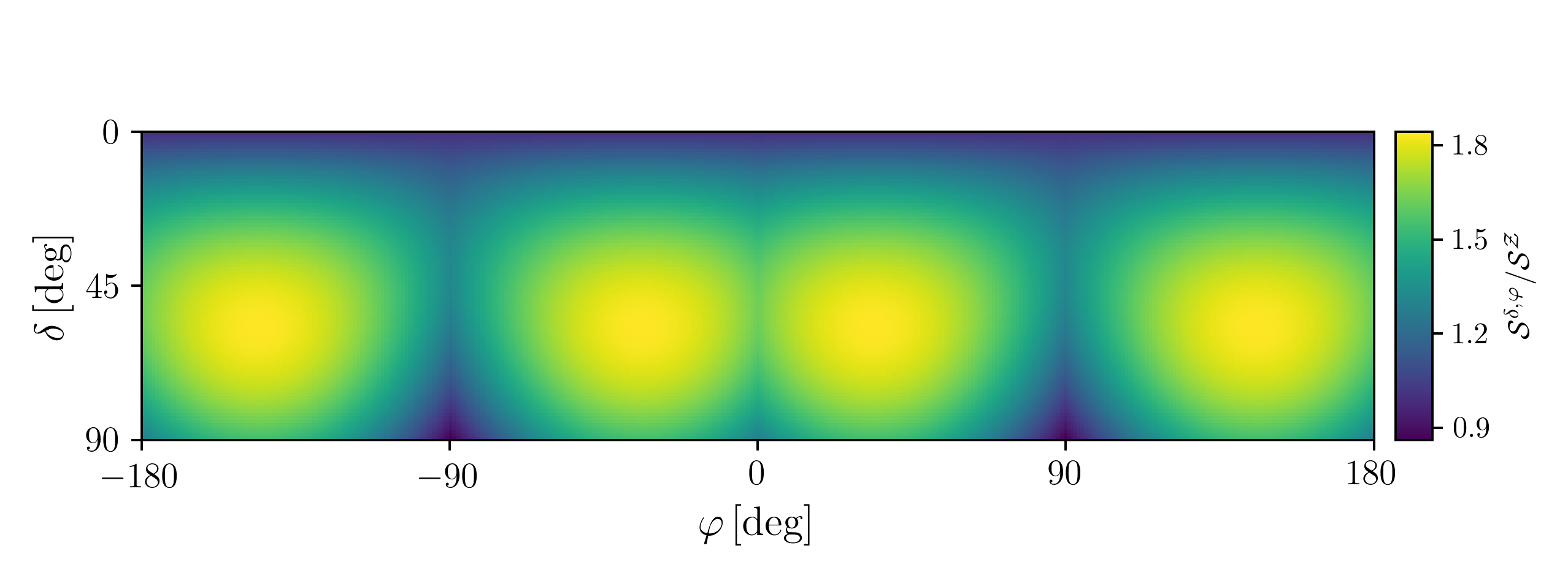} 		\centering 
		\caption{Significance of the signal modulation $\pazocal{S}^{\delta,\varphi}[\sigma]$ for a head-on axion event. Being $\delta$ the inclination angle measured from a fixed zenith direction ($\pazocal{Z}$), $\varphi$ the azimuthal angle of its orthogonal projection on a north-west ($\pazocal{N-W}$) reference plane, measured from the north ($\pazocal{N}$) fixed reference. Full telescope rotation normalized to the signal modulation significance of a zenith pointing experiment ($\pazocal{S}^\pazocal{Z}[\sigma]$). Observed in the Canary Islands during the first day of the year. }
		\label{fig_10}
	\end{figure}

The experiment in Fig. \ref{fig_2} is shown pointing $\pazocal{Z}$. Altazimuth coordinates can be used to express a telescope rotation. The pointing tensor is $\mathord{\mathrm{\hat{e}}}_{\delta,\varphi}=(\mathord{\mathrm{sin}\:\!\delta \:\! \mathord{\mathrm{cos}\:\!\varphi}}, \mathord{\mathrm{sin}\:\!\delta \:\! \mathord{\mathrm{sin}\:\!\varphi}}, \mathord{\mathrm{cos}\:\!\delta} )$, where $\delta$ and $\varphi$ are the declination angle and the azimuthal angle in lab spherical coordinates, respectively. The transformation $c_i^{\delta,\varphi}=\mathord{\mathrm{\hat{e}}}^{\delta,\varphi} \, c_i^{\pazocal{N,W,Z}}$ is obtained from multiplying by $\mathord{\mathrm{\hat{e}}}^{\delta,\varphi}$ in Eq. \ref{Eq.11}. Aimed to illustrate the features of a telescopic scan for axion we include Fig. \ref{fig_10}. Here, the detection significance\footnote{As defined in ref. \cite{Knirck_2018}.} of a telescope measuring a head-on crossing event is compared with the case of a stationary dielectric haloscope, with directionality given only by its geometry \cite{Millar_2017_2}. The preferred direction varies with lab coordinates ($\lambda_{\mathord{\mathrm{lab}}}, \phi_{\mathord{\mathrm{lab}}}$) and during a year, and can be tracked in order to enhance the signal modulation. Fig. \ref{fig_10} shows that a telescopic scan is up to two times more significant. Interestingly, this result is independent from instrumental sensitivity, axion model and mass, substructure density, velocity and dispersion. We also estimate that our experiment presents a potential to detect head-on events with a QCD axion density about several tens the local DM density ($\rho_{DM}$) and a velocity dispersion $\pazocal{O}(10)$ km$\,\!$s$^{-1}$ with a $3\sigma$ significance over a few day integration time by measuring its daily modulation. A denser occupation number, a more favourable relative velocity between the lab and the flow or a lower dispersion would result in a higher confidence level.

\section{Summary and conclusions}
\label{sec:discussion}
Axion detection would be one of the most important moments of the entire history of Science. Given the fruitless searches of other dark matter candidates \cite{Schumann_2019} and the discovery in July 
2012 of a Higgs-like boson (\cite{Aad:2012tfa}, \cite{Chatrchyan:2012xdj}), the first fundamental 
particle of a scalar nature,\footnote{Notice that axion is a pseudo-scalar particle.} there is a
 renewed effort in the scientific community to  characterize experimentally and find the axion in the 
parameter space in which this can simultaneously solve the mystery of dark-matter and the problem of CP symmetry 
of the strong interaction. This energy space is weak and requires ultra sensitive detectors. In order to 
face this challenge, the haloscope was proposed \cite{PhysRevLett.51.1415}.  

In this article, we have presented a pioneering halo-telescope of high sensitivity and remarkable simplicity
able to probe the 25--250 $\mu$eV mass range. The proposed set-up has the advantage of simultaneously exploring axion-like particles and dark photon sectors during scanning. A highlighted novelty of this experiment is the possibility of telescopic scanning. The apparatus 
is sensitive from low-velocity up to relativistic dark matter particles. The relevance on directionality and the exploration of non-virialized particles (astroparticles) is remarkable. The haloscope is the most sensitive instrument to date for the detection of relic axions, 
ALPs and dark photons. However, in the case that cold dark matter is distributed in the form of substructures 
(miniclusters, minihalos, etc.), and that Earth is placed  within a low density region 
($\mathord{\mathrm{\rho_{DM}<<300 \, MeV\,cm^{-3}}}$), haloscopes present a weakness 
since they depend on the local halo density. In this case, dark matter astroparticle scanning (assuming a significant
 particles flux) using telescopes and helioscopes, or laboratory experiments in which the axion pump is 
artificially generated, could be the only direct detection strategy. Note that heterogeneous populations are additive if they have a similar dynamic mass, resulting in an improved signal-to-noise ratio. On the other hand, mounting the detector on an altazimuth platform allows functionality to be added to the experiment at an assumable cost.
\textit{Axioastronomy} could probe several theories and hypotheses.\footnote{See section \ref{sec:Directional} 
for an extended discussion.}

Another remarkable advantage of the experiment is that it has been designed prioritizing the use of state-of-the-art (commercial)
technology. 
The lack of the need for R\&D reduces the cost in hardware and benefits from faster manufacturing and easier 
maintenance. On the other hand, the decision to use an (MRI type) commercial multicoil
 magnet, although this limits the size and the magnetic inductance that is possible to incorporate into 
the experiment, results in economy and reduced global fabrication time. We estimate that the cost of the experiment is reduced by 
an order of magnitude thanks to this strategy. That does not mean that the experiment could not benefit from the incorporation of incipient technologies.\footnote{The MSA-based radiometer included in Fig. \ref{fig_9}. Previously used in low frequency magnetized experiments, this technology might be feasible at higher frequencies.}

The experiment presents a potential to explore the QCD axion sector over the entire 6--60 GHz band sequentially, in sub-bands of the order of 100 MHz wide, each taking from a few 
weeks to several months of time for scanning, depending on the observational strategy and scan mode. In case of marginal detection, more time can be spent scanning the 
sub-band of interest. The reconfiguration needed for scanning the 6--60 GHz band is small and consists in grid resizing and
the replacement of the microwave receivers.\footnote{Standard radiometers have a typical bandwidth factor 1.4:1.} 

Finally, graviton also mix with ordinary photons in the presence of external static magnetic fields (\cite{Gertsenshtein}, \cite{Cembranos:2018jxo}). Cosmology predicts a stochastic gravitational wave background (GWB) generated by a number of mechanisms in the Universe.\footnote{The coalescence of primordial black hole (PBH) binaries may produce a stochastic GWB component, while the evaporation of low-mass PBHs would produce high-frequency (MHz-GHz) GWs. These PBHs may be a DM component (\cite{BisnovatyiKogan:2004bk}, \cite{Marck:308812}). \textit{Branes} oscillation may be a different GWB component in the band $10^8-10^{14}$ Hz (\cite{Servin:2003cf}, \cite{Clarkson}). The nucleosynthesis upper limit is $\mathord{\mathrm{\Omega}}_{GW}\gtrsim10^{-5}$ \cite{Maggiore:1999vm}.} Several experiments have been created to constraint the amplitude of the GWB in narrow bands, presenting a limited sensitivity (\cite{Chou:2016hbb}, \cite{Cruise:2006zt}, \cite{Akutsu:2008qv}, \cite{Nishizawa:2007tn}, \cite{Ito:2019wcb}). The analysis carried out in \cite{Ejlli:2019bqj} using data from the LSW experiments ALPS and OSQAR in (\cite{OSQAR}, \cite{Ehret:2010mh}) and the CAST helioscope \cite{Anastassopoulos:2017ftl} establishes exclusion regions for the GWB of the order of $h_c \gtrsim 10^{-24}-10^{-27}$ in several sub-bands in the THz and optic range, where $h_c$ represents the (dimensionless) minimum detectable characteristic amplitude of a stochastic, isotropic, stationary and unpolarized GWB. We explore the possibility that DALI might provide similar bounds for the GWB in the 6-60 GHz band.

\section*{Acknowledgements}
I would like to thank J. Mart\'in Camalich and R. Rebolo for discussions. Thanks K. Zioutas, A. Zhitnitsky, J. Redondo, A. Sedrakian, E. Hern\'andez and F. Gracia for comments.




\end{document}